\documentstyle [12pt, epsfig]{article}

\begin{document}

\centerline{\bf ABOUT POSITION MEASUREMENTS}
\centerline{\bf WHICH DO NOT SHOW THE BOHMIAN
PARTICLE POSITION}
\vskip 1.6cm

\centerline{\bf Yakir Aharonov$^{a,b}$ and Lev Vaidman$^a$}

\vskip 1cm

\centerline{\it $^a$  School of Physics and Astronomy}
\centerline{\it Raymond and Beverly Sackler Faculty of Exact Sciences}
\centerline{\it Tel-Aviv University, Tel-Aviv, 69978 ISRAEL}

\vskip .5cm

\centerline{\it $^b$ Physics Department, University of South Carolina}
\centerline{\it Columbia, South Carolina 29208, U.S.A.}

\vskip 1.6cm
This work is inspired by our discussions with David Bohm in different
periods in Tel-Aviv, London, and South Carolina. We were excited by the
results we obtained in the framework of two-state vector formalism about
weak measurements (Aharonov and Vaidman 1990) and we were trying to
understand these results using Bohm's causal interpretation (Bohm
1952). Both Bohm's theory and the two-state vector formalism yield the
same predictions for the results of experiments as the standard quantum
theory.  Therefore, clearly, there cannot be a technical contradiction
between these two approaches. However, it is a legitimate question to ask
how close or how different are the concepts of the two theories. Friction
between the basic concepts might lead to a direction for the  modification of
quantum theory;  the search for a useful modification was the goal of David
Bohm.

I (L.V.) have a vivid memory of an excursion day after the conference in
South Carolina honoring 30 years of the Aharonov-Bohm effect when I, David
and Sarah Bohm were riding in a carriage in the old streets of
Charleston. I remember discussing with David my passion for  the many-worlds
interpretation (Everett 1957), and then, what I find most attractive in the
Bohm  theory: that
it is unambiguous, deterministic and complete. I see it as the best
candidate for the final theory of the world for a physicist (such as Y.A.)
who does not want to accept the existence of many worlds. David said to me
that what I like in his theory has not much value for him. He had a strong
belief that a man cannot find the final theory of the world. All that we
can  do is to look for a better and better approximation to the correct
theory which is intrinsically unattainable to us. His vision was that the
causal interpretation should suggest a way for generalization to the
next-level theory which also, by no means, will be the final theory of the
world.

We believe that the observations we make in this work are somewhat
disturbing for a physicist who wants to see in the causal
interpretation the final word about the world, but they are not a real
threat for the theory from the perspective of Bohm.  Apart from
discussing weak measurements we will consider recently proposed
``delayed observation'' measurements (Englert et al. 1992) which
exhibit similar features. Note, however, that the difficulties we see
follow mostly from a particular approach to the Bohm theory we adopt,
in which only the Bohmian particle corresponds to the ``reality''
which we experience, while the wave function is just a pilot wave
which governs the motion of the particle.  By the position of the
``Bohmian particle'' we mean what is frequently called actual position
of the particle in the Bohm theory.  For a composite system consisting
of many particles we shall, somewhat abusing language, refer to the
point in configuration space formed from the the coordinates of all
the Bohmian particles as the ``Bohmian particle.''

We consider the following two principles desirable for a causal
interpretation. In Bohmian mechanics in most cases they are indeed
valid. However, as we will show, there are situations in which they are
not.

I) {\it A procedure which we usually consider as a good measurement of position
should yield the position of the Bohmian particle.}

II) {\it An empty wave (the one without the Bohmian particle inside) should not
yield observable effects on other particles.}

\noindent
The motivation for the first principle is clear. The second principle we
find desirable because otherwise the Bohmian picture becomes very
complex. In the Bohm theory there is no collapse of the wave function, so
the total wave incorporates all the complexity of the many worlds of the
Everett interpretation. We hoped that the Bohm theory could avoid it.  The
Bohm theory, as we understand it, says that when we observe a person whose
Schr\"odinger wave is split into a superposition of two macroscopically
different waves, we actually see only the component of the superposition
which has the Bohmian particle inside. We thought that
 this is true also when we ``observe''  a single particle -- by means of
 any type of interaction that we might speak of as detecting it.

Let us start with the following example. The wave function of the particle
consists of two identical wave packets running in opposite directions, see
Fig. 1. For simplicity we will \phantom{llllllllll}

\epsfysize=8 cm
 \centerline{\epsfbox{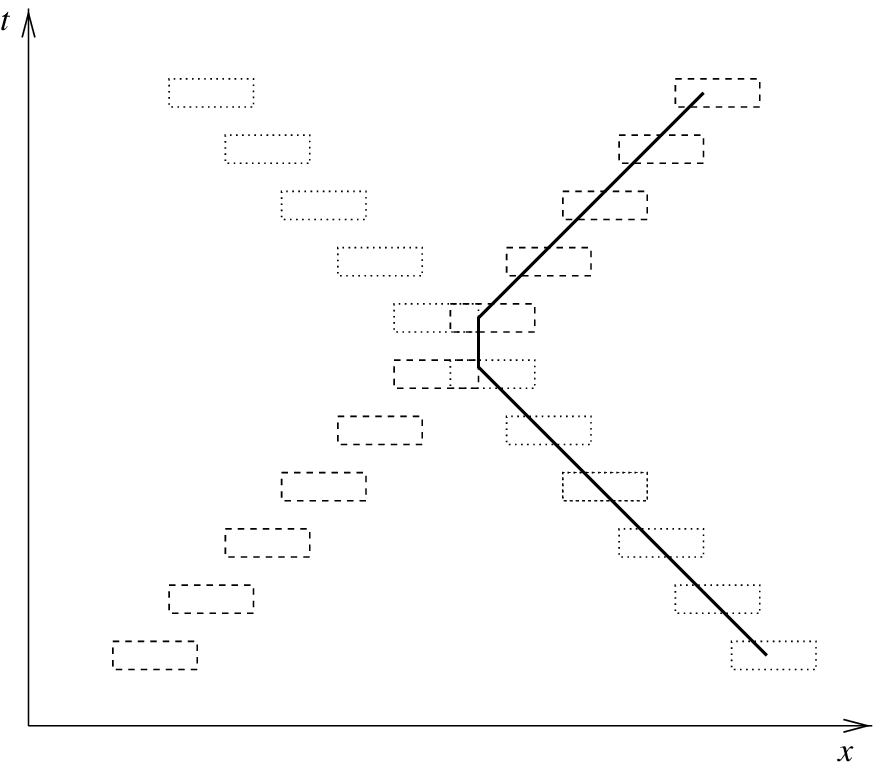}}
{\small {\bf Figure 1}: {\footnotesize
 Space-time diagram for the wave and the Bohmian
particle. The initial state is the superposition of two
identical wave packets moving one towards the other. The Bohmian
particle located initially in the right wave packet. Contrary to the
naive expectation the Bohmian particle makes a turn in the region
where no (Hamiltonian type) interaction takes place.
}}

\noindent
 consider a rectangular shape and assume that
the spread of the wave packets during the process can be neglected.  In the
beginning the Bohmian particle is inside the right wave packet. The Bohm
theory yields an extremely simple prescription for finding the evolution of
the system. Until the overlap, the Bohmian particle runs together with the
right wave packet. At the moment it reaches the area of the overlap, it
stops (the currents of the two wave packets cancel each other in the
calculation of the Bohmian particle velocity). Since at that moment the
left wave packet has its whole length to go over the particle, while the
right wave packet has only a part of its length, the particle will end up
inside the left wave packet and will run with it to the right.

 One of us (L.V.) used to view the Bohm interpretation as the most elegant
way of pointing  out one of the many worlds of the Everett interpretation as
``real''. This example shows that the Bohmian world might be different from
any of the Everett worlds. Indeed, no turns take place in Everett worlds in
this example.

Now let us add a robust position measurement on the left side of the wave
packet (the empty wave); see Fig. 2. By robust measurement we mean a von
Neumann type interaction between the particle and the pointer variable of
the measuring device. We can model it by the interaction Hamiltonian
 $$
 H =
g(t) P \Pi_{V},\eqno(1)
 $$
 where $P$ is the conjugate momentum to the pointer variable $Q$ which is,
say, the spatial coordinate of the pointer.  $\Pi_V$ is a projection
operator on the volume $V$. The left wave packet of the particle passes the
volume $V$ during a certain time $t$. The coupling constant $g$ is chosen in
such a way that during that time the wave function of the pointer is
shifted by a distance much larger than its spread.  We consider the initial
state of the particle and the position of its Bohm coordinate to be the
same as in the previous case. Of course, initially the Bohmian particle of
the pointer is somewhere inside its initial wave packet. It is very easy to
see the evolution, according to the causal interpretation, also in this
situation. The wave function of the particle and the system become
entangled; the right wave packet is entangled with no shift of the pointer
while the left wave packet is entangled with the shifted wave packet of the
pointer. The Bohmian particle which started inside the right
wave packet runs with it all the way to the left, and the Bohmian particle
of the pointer does not move. Indeed, when the wave packets of the particle
overlap, there is no overlap for the wave function of the composite system
(particle and the pointer) and therefore the left wave packet does not
change the motion of the Bohmian particle moving inside the right wave
packet; see Fig. 2a. (If, instead, initially the Bohmian
particle was inside the left wave packet, then it will run with this wave
packet  all the
way to the right, while the Bohmian particle  of the pointer will move inside
the shifted wave packet of the pointer outside its initial location; see
Fig. 2b.)

In this example we can see the two principles at work. The position
measurement shows that the particle is on the left if and only if the
Bohmian particle was on the left. And, also, observable action on the
measuring device occurs  only  when the Bohmian particle is
at the location of the measuring device.

Let us turn now to an example in which there is some difficulty with causal
interpretation.  We will consider again the same situation but, instead of
a robust measurement, we will discuss ``weak measurement". This is a
standard measurement with a weakened coupling. The difficulty appears when
we consider the pre- and post-selected ensemble. The particles are all
pre-selected in the initial state which is the superposition of the two
wave packets as

\epsfysize=16 cm
\centerline{ \epsfbox{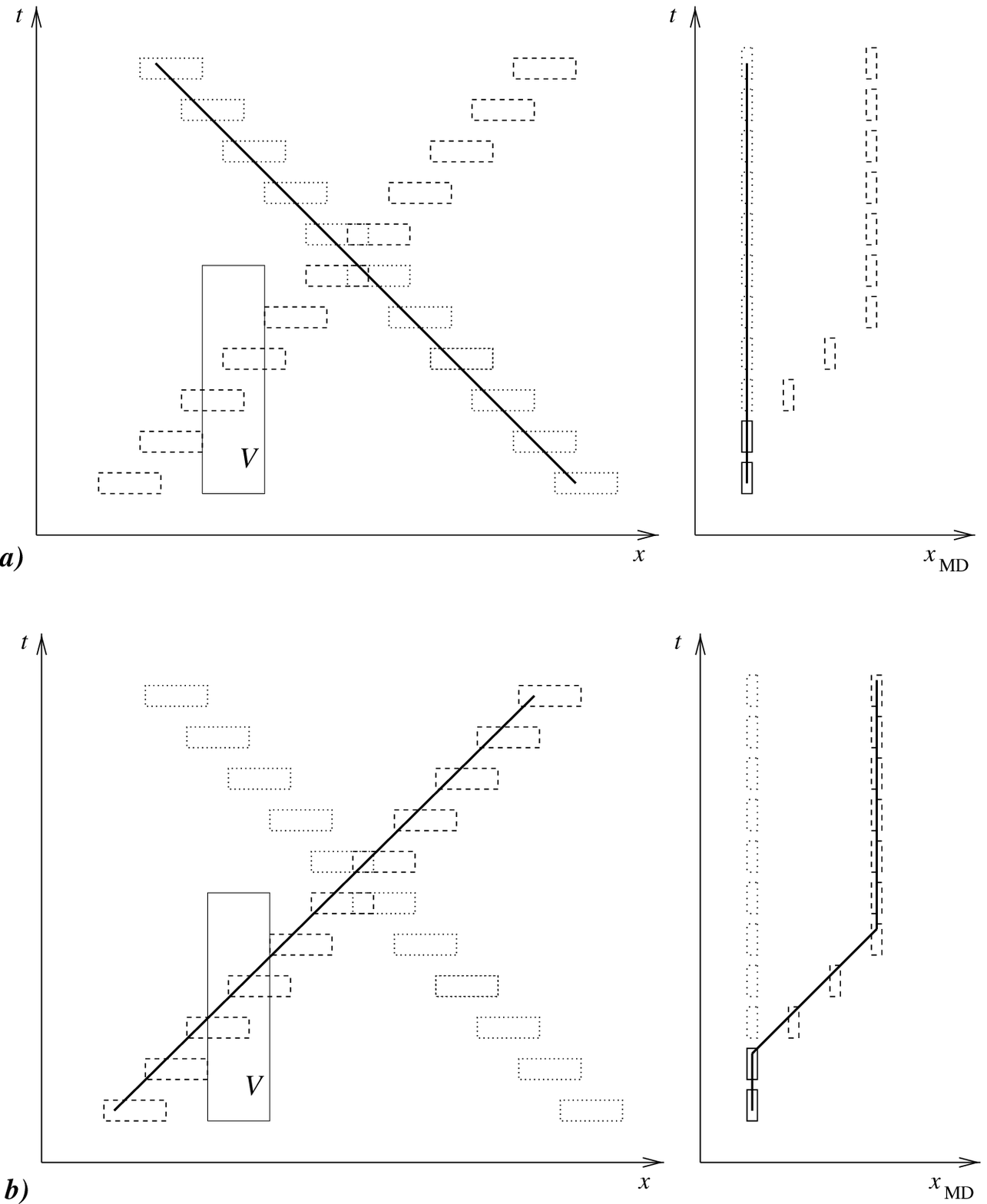}}
{\small {\bf Figure 2}: {\footnotesize
 Space-time diagram for the wave and the
Bohmian particle and the wave and the Bohmian particle of the
measuring device. {\bf a).} The Bohmian particle starts on the right
and does not enter the interaction region $V$. The Bohmian particle of
the pointer of the measuring device does not move. {\bf b).} The
Bohmian particle starts on the left, passes through the interaction
region $V$ and the Bohmian particle of the pointer moves, thus showing
that the particle was on the left side.
}}

\vspace{.5cm}

\noindent
 described above. At the end we observe the location of the
particle and consider only the cases in which the particle is found in the
right side. Contrary to the example above, no {\it a priori} assumption is
made about the initial Bohm coordinates. Since the final measurement is
considered to be robust, we assume that the final Bohm coordinate is in the
right side too.

For simplicity, instead of taking the usual ``weak measurement'', we
will consider now a simple model which, in this case, exhibits similar
features. We will assume that the wave function of the pointer has a
rectangular form and that its spread during the process can be
neglected. The coupling constant $g(t)$ is taken to be very small such
that the shift caused by the particle passing through the volume $V$
is equal to 10\% of the width of the wave packet; see Fig. 3.  In this
situation one measurement yields usually (in $90\%$ of the cases) no
information, but the outcomes of a number of results obtained on a
large ensemble of identical\break

\epsfysize=16 cm
  \centerline{\epsfbox{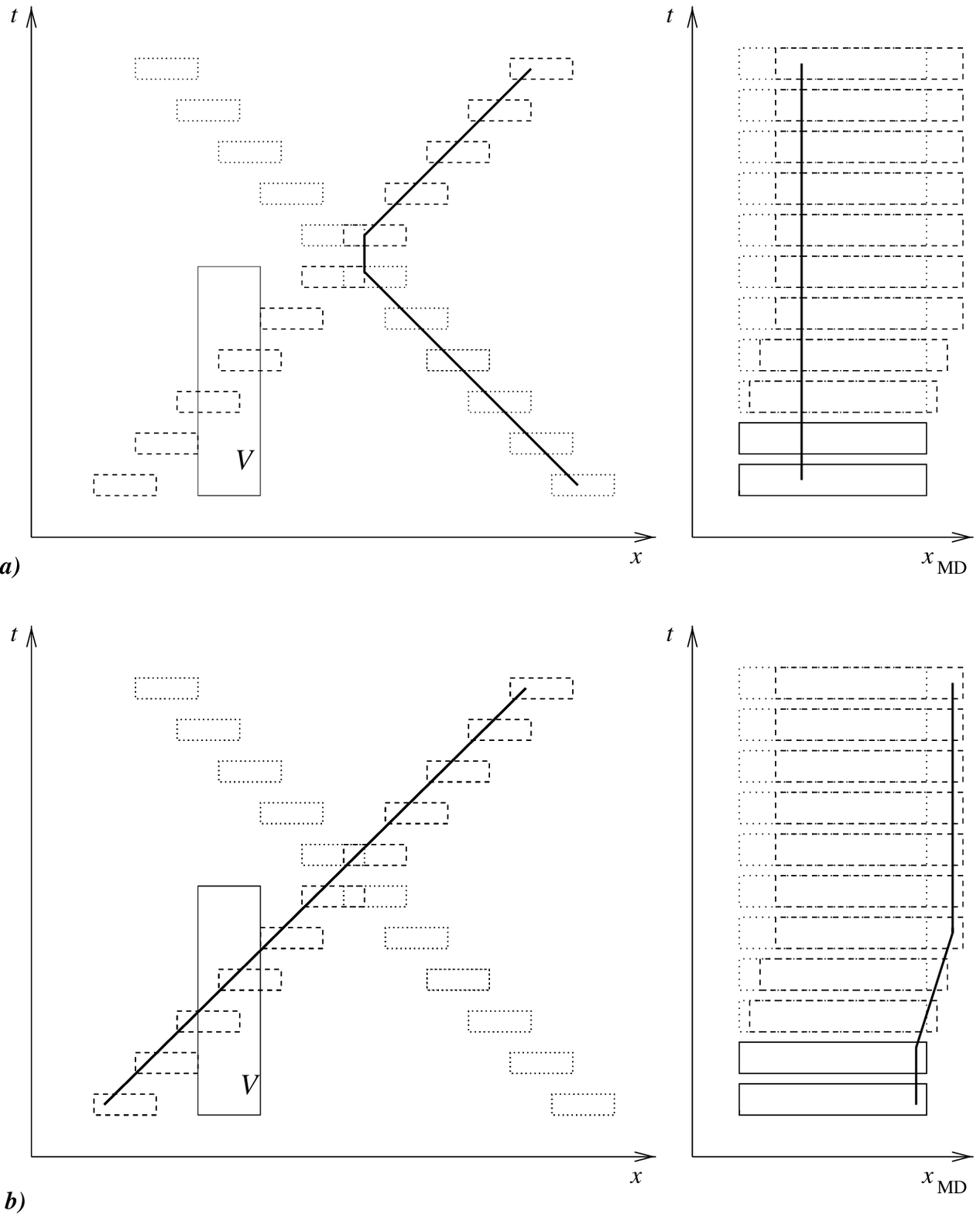}}
{\small {\bf Figure 3}: {\footnotesize
 Space-time diagram for the ``weak
measurement''. Only the cases when the particle ends up on the right
are considered.  {\bf a).} The Bohmian particle starts on the right
and the Bohmian particle of the pointer starts anywhere in the right
90\% of the wave. {\bf b).}~The Bohmian particle starts on the left
and the Bohmian particle of the pointer starts anywhere in the right
10\% of the wave.  The final distribution of the Bohm coordinates of
the measuring device corresponds (according to the usual, pre-selected
only, situation) to the particles which were initially on the left.
Indeed, the Bohm coordinate of the pointer cannot be found at the end
in the first (left) 10\% of the initial wave. Nevertheless, 90\% of
the Bohmian particles started on the right.
}}

\noindent
 pre- and post-selected systems will yield
a clear answer: the particle is on the left. Indeed, the statistical outcomes
are identical to those
obtained from measurements on a pre-selected ensemble with a particle
placed on the left side.  In order to fulfill the first principle we need
that in such a situation we can claim with high probability that the
Bohmian particle was also on the left. This is, however, not so.

Assuming uniform distribution of the Bohmian particles we can easily see
that in $90\%$ of the cases, in spite of the results of the measurements,
the Bohmian particle was not there. Indeed, if the Bohmian particle of the
pointer was anywhere in the last $90\%$ of its wave packet (i.e.  not close
to the beginning edge) and the Bohm coordinate of the particle was on the
right, then it will behave in the same way as in the first example: the
Bohmian particle will move together with the right wave packet, stops when
it enters the region of the overlap of the two wave packets until it is
taken to the right by the left wave packet; see Fig. 3a.  For Bohmian
particle starting on the left side the situation is similar. In $90\%$
cases, when the pointer Bohmian particle is in the beginning part of the
wave function, the Bohmian particle makes the turn and, only when the
Bohm's pointer coordinate is in the last 10\% percent of the wave function,
does it go straight to the right; see Fig. 3b.  Thus, the $90\%$ of the
Bohmian particles which ended at the right side, started from the right,
and only $10\%$ started from the left. The apparent difficulty is that the
outcomes of the measurements, if interpreted in a natural way, correspond
to all particles being on the left side, while only $10\%$ of the Bohmian
particles were there.

It is important to emphasize that we have difficulty only with the first
principle. The change in the measuring device in all cases can be seen as a
direct influence of the nonempty waves. Although we had just $10\%$ of
particles on the left side, for all of them  the  Bohmian particle of the
pointer was at
the far end, such that after the shift due to the interaction, they fill
the missing part of the spatial distribution of the pointer relative to the
distribution corresponding to no particle on the left.

The difficulty which we have in this example can be seen only on an
ensemble of pre- and post-selected systems. We may try to consider the
whole ensemble of $N$ particles as a single system, but then the difficulty
will not appear.  Indeed, in this case we have to consider a single
measuring device. If we take the same measuring device which is coupled,
one after the other, to all particles, then the picture becomes different.
Straightforward analysis shows that for $N \gg 10$ most of Bohmian
particles start from the left and pass the volume $V$.

In order to explain our next example, which is a weak measurement performed
on a single (pre- and post- selected) system, we will start with a brief
review of the two-state vector formalism. In 1964 Aharonov, Bergmann and
Lebowitz considered measurements performed on a quantum system between two
other measurements, results of which were given. They proposed describing
the quantum system between two measurements by using two states: the usual
one, evolving towards the future from the time of the first measurement,
and a second state evolving backwards in time, from the time of the second
measurement.  If a system has been prepared at time $t_1$ in a state
$|\Psi_1\rangle$ and is found at time $t_2$ in a state $|\Psi_2\rangle$,
then at time $t$, $t_1<t<t_2$, the system is described by $$
\langle \Psi_2 | e^{i\int_{t_2}^{t} H dt}
{\rm ~~~ and~~~}
e^{-i\int_{t_1}^{t} H dt} |\Psi_1\rangle . \eqno(2)
 $$
 For simplicity,  we shall consider the free Hamiltonian to
be zero; then, the system at time $t$ is described by the two states
 $
\langle \Psi_2 |
$ and $ |\Psi_1\rangle $. In order to obtain such a system, we prepare an
ensemble of systems in the state $ |\Psi_1\rangle$, perform a measurement
of the desired variable using separate measuring devices for each system in
the ensemble, and perform the post-selection measurement. If the outcome of
the post-selection was not the desired result, we discard the system and
the corresponding measuring device. We look only at measuring devices
corresponding to the systems post-selected in the state $\langle \Psi_2 |$.

 The basic concept of the two-state approach, the weak value of a physical
variable $A$ in the time interval between pre-selection of the state $|
\Psi_1 \rangle$ and post-selection of the state $ | \Psi_2 \rangle$ is
given by (Aharonov and Vaidman 1990) $$ A_w \equiv {{\langle \Psi_2 | A |
\Psi_1 \rangle}
\over {\langle \Psi_2 |\Psi_1 \rangle}} ~~~~.
\eqno(3)
$$
 Let us show briefly how weak values emerge from a measuring procedure
with a sufficiently weak coupling.  We consider a sequence of measurements:
a pre-selection of $|\Psi_{1} \rangle$, a (weak) measurement interaction of
the form of Eq. (1), and a post-selection measurement finding the state
$|\Psi_2 \rangle$.  The state of the measuring device (which was initially
in a Gaussian state) after this sequence is given (up to normalization) by
$$
\Phi (Q) = \langle \Psi_2 \vert
e^{-iPA}
\vert \Psi_1 \rangle e^{ -{{Q ^2} /{2\Delta ^2}}} ~~~~.
\eqno(4)
$$
 In the
$P$-representation we can rewrite it  as
$$
\tilde \Phi (P) =   \langle \Psi_2
\vert \Psi_1 \rangle ~ e^{-i {A_w} P} ~
e^{-{{\Delta}^2 {P^2}} /{2}} ~ + ~\langle \Psi_2 \vert
 \Psi_1 \rangle  \sum_{n=2}^\infty {{(iP)^n}\over{n!}}
[(A^n)_w - (A_w)^n]   e^{ -{{\Delta ^2 P^2}} /{2}}~.\eqno(5)
$$
\noindent
If $\Delta$ is sufficiently large,  we
can neglect the second term of (5) when we Fourier transform
 back to the  $Q$-representation.  Large $\Delta$
corresponds to weak measurement in the sense that the
interaction Hamiltonian
(1) is small.  Thus, in the limit of weak measurement, the final state
of the measuring device (in the $Q$-representation) is
$$
\Phi  (Q) = e^{ -{{(Q - A_w)^2} /{2\Delta
^2}}}~~~~. \eqno(6)
$$
This state represents a measuring device pointing to the weak value, $A_w$.

Although we have showed this result for a specific von Neumann model of
measurements, the result is completely general: any coupling of a pre- and
post-selected system to a variable $A$, provided the coupling is
sufficiently weak, results in effective coupling to $A_w$. Since $\Delta$
has to be large, the  weak
coupling between a single system and the measuring device will not, in most
cases, lead to a distinguishable shift of the pointer variable, but
collecting the results of measurements on an ensemble of pre- and
post-selected systems will yield the weak values of a measured variable to
any desired precision.

As an example, consider a spin-1/2 particle prepared with the spin
state pointing in the $x$ direction and found later with the spin in
the $y$ direction.  We consider weak measurement of the spin component
in the ${\xi}$ direction which is the bisector of $\hat{x}$ and
$\hat{y}$, i.e.,
$
\sigma_\xi =   (\sigma_x + \sigma_y)/\sqrt 2
$.
 Thus,
${|}\Psi_1 \rangle =|{\uparrow_x} \rangle$,
$|\Psi_2 \rangle =|{\uparrow_y} \rangle$, and  the weak
value of $\sigma_\xi$  in this case is:
$$
(\sigma_\xi)_w =
 {{\langle{\uparrow_y} |\sigma_\xi  |{\uparrow_x} \rangle}
\over {\langle{\uparrow_y} |{\uparrow_x} \rangle}} =
 {1\over\sqrt 2}{{\langle{\uparrow_y} | (\sigma_x + \sigma_y) |{\uparrow_x}
\rangle}
\over {\langle{\uparrow_y} |{\uparrow_x} \rangle}} = \sqrt 2 ~~.
\eqno(7)
$$
This value is, of course, ``forbidden'' in the standard interpretation
where a spin component can obtain the (eigen)values $\pm1$ only.
 If the angle between the
spin directions of the initial and final states is close to $180^0$, then
the weak measurement in the direction of the bisector can be even much larger
(Aharonov et al. 1988).

Discussing spin-measurements in the framework of the Bohm theory requires
 additional care. The most elegant way to analyze spin in causal
 interpretation is not to attribute to it some kind of a Bohm pointer
 (e.g. Dewdney et al.  1988), but to consider the spin as a property of the
 quantum wave. Then the probability one-half in the result of a spin $z$
 measurement of a spin prepared in, say, the $\sigma_x =1$ state
 corresponds to uniform distribution of the Bohmian particles inside the
 wave packet. If the Bohmian particle is in the upper half of the wave, the
 Stern-Gerlach measurement will end up with the particle going up;
 otherwise it will go down.  This is a manifestation of {\it contextuality}
 of quantum measurement: the fact that the particle goes up is interpreted
 as spin ``up'' or spin ``down'' depending on the direction of the gradient
 of the magnetic field in the Stern-Gerlach device (see, for example,
 Albert 1992, pp.153-55).

 Bohm's picture nicely explains the peculiar result of a weak measurement
of the spin component. Due to the coupling of the weak measurement, the
Bohm coordinates of the pointer variable (which is the coordinate of the
particle) are not shifted by the ``forbidden'' value
$(\sigma_{\xi})_w$. The interaction causes correlations such that the
post-selection measurement will pick the Bohmian particle from the
appropriate region.  In the case of a large value of $(\sigma_{\xi})_w$, the
post-selected Bohmian particles are coming from the ``tail'' of the wave
function. We will not present here the derivation of this result, but we
will give a hint how it works. Assume that the initial state of the
particle is a Gaussian with the spread $\Delta $. Then, to calculate the
motion of the Bohmian particles in the post-selection measurement we have to
consider a mixture of two, almost identical, slightly shifted
Gaussians. The particles in a certain region go towards a certain
(post-selection) direction if the ``weight'' of one packet is significantly
larger than that of the other. But this, for very small shift, happens far
away in the tail: $e^{{-(x-\delta)^2}\over
\Delta^2}/ {e^{{-x^2}\over \Delta^2}}  \gg 1 ~~{\rm for }~~ x \gg \Delta^2
/ \delta $.

The example we want to consider here is less interesting for practical
applications, but more striking conceptually. It is a weak measurement
performed on a single system with a particular (very rare) post-selection
which does lead to a distinguishable shift of the pointer.  Such was the
example considered in the first work on weak measurements (Aharonov et
al. 1987). In this work a single system of a large spin $N$ is
considered. The system is pre-selected in the state $|\Psi_1\rangle = |S_x
{=} N \rangle$ and post-selected in the state $|\Psi_2\rangle = |S_y {=} N
\rangle$.  At an intermediate time the spin component $S_\xi$ is weakly
measured and again the ``forbidden" value $\sqrt 2 N$ is obtained. The
uncertainty has to be only slightly larger than $\sqrt N$.  The probability
distribution of the results is centered around $\sqrt 2 N$, and for large
$N$ it lies clearly outside the range of the eigenvalues, $(-N, N)$. More
generally, we can consider pre-selection and post-selection in the
directions such that the angle with the bisector $\hat \xi$ is
$\theta$. Then, the weak value of $S_\xi$ is $({S_\xi})_w = {N\over
\cos\theta}$ and it can be much larger than $N$.

Our purpose here is to consider a position measurement performed on a
single system. This is, however, different from the ``position''
measurements we discussed above. In previous examples we measured a
projection operator on a certain volume. Now we consider a measurement
of the operator of position. Our example is similar to the spin-$N$
particle experiment discussed above. We consider a massive particle
which was prepared in a state $|\Psi_1\rangle$ and was found at time
$t_2$ in a state $|\Psi_2\rangle$.  The states $|\Psi_1\rangle$ and
$|\Psi_2\rangle$ are the superpositions of very well localized wave
packets around locations between -1 and 1. The states, up to
normalization, are

{}~~~~~$
|\Psi_1\rangle = \sum_{i=0}^N |x = {{N-2i}\over N}\rangle, ~~~~~~~~~~~
|\Psi_2\rangle =\sum_{i=0}^N  c_i
|x = {{N-2i}\over N}\rangle, ~~~~~~~
$\hfill (8)

\vskip .2cm
\noindent
where $c_i = {{(-\tan^{2} \theta/2)^i}\over {i!(N-i)!}}$.  In this case the
weak measurement (the standard measurement of position with  precision
of the order $\sqrt N$) yields a number around the weak value of $x$ which
is $ x_w = {1\over {\cos\theta}} $.  Mathematically, this is identical to
the superposition of small forces which is equivalent to a large force, as
discussed by Aharonov, Anandan, Popescu and Vaidman (1990). The details can
be found there.

If we choose $\theta$ and $N$ such that $\sqrt N \ll {1\over {\cos\theta}}$
we may view this experiment as a good measurement of position: we can
repeat the procedure several times (including the pre-selection, the
position measurement and the post-selection) and we will obtain values
around $x_w$ with relatively small statistical spread. Nevertheless, the
outcome of this kind of position measurement has no relation to the Bohmian
particle position; the latter can have only values between $-1$ and 1. It
is not that there is any real contradiction, or that the causal
interpretation cannot explain the outcome of the measurement. If we
consider the Bohmian particle position of the pointer variable, it will be
around the value $x_w$. Again, it was not shifted by the coupling of weak
measurement, but ``picked'' from the tail in the post-selection process,
and this is due to the peculiar wave function correlation created by the whole
process. The unfortunate feature is that again, the first principle is not
fulfilled in this specific case: the procedure which is very close to the
usual position measurement yields consistently results which have no
relation to the Bohmian particle position.

Another example where the Bohmian particle position does not help to understand
the
result of measurement is the {\it protective measurement} (Aharonov and
Vaidman 1993) of the Schr\"odinger wave function. If the only ``reality''
is the Bohmian particle position and the Schr\"odinger wave is just a pilot
wave which governs the motion of the particle, then the wave should not be
observable directly. Let us briefly show how, in certain cases, we can
observe the complete Schr\"odinger wave of a single particle, in spite of
the fact that the Bohmian particle position represents almost no features of
the wave.

As an example of a simple protective measurement, let us consider a
particle in a discrete nondegenerate energy eigenstate $\Psi (x)$. The
standard von Neumann procedure for measuring the value of the projection
operator, $\Pi_V$, on the volume $V$ involves an interaction Hamiltonian
(1).  In the protective
measurements $g(t) = 1/T$ for most of the time $T$ and it goes to zero
gradually before and after the period $T$.  For $g(t)$ smooth enough we
obtain an adiabatic process in which the particle cannot make a transition
from one energy eigenstate to another, and, in the limit $T \rightarrow
\infty$, the interaction Hamiltonian does not change the energy eigenstate.
For any given value of $P$, the energy of the eigenstate shifts by an
infinitesimal amount given by first order perturbation theory: $\delta E =
\langle H_{int} \rangle = {
\langle
\Pi_V \rangle P}/ T.$
The corresponding time evolution $ e^{-i P \langle \Pi_V \rangle} $
shifts the pointer by the average value $\langle \Pi_V
\rangle $.
By measuring the averages of a sufficiently large number of projection
operators of different regions, the density of the full Schr\"odinger wave
can be reconstructed to any desired precision.

Again, we would like naively to think that the projection operator $\Pi_V$
should yield 1, if the Bohmian particle is inside $V$ and zero if it is
outside $V$. The measurement, however yields $|\Psi(x)|^2$ irrespectively
of where the Bohmian particle is. If the system is in an energy eigenstate,
then the Bohmian particle does not move, so, in fact, for measurements of
almost all projection operators the Bohmian particle is outside the volume
during the whole period of the measurement, and for one projection operator
it is always inside. Thus, the second principle is broken here: the
measuring devices measuring $|\Psi(x)|^2$ in the ``empty'' volumes yield
non-null outcomes.

 More precisely, the measuring interaction changes the state of the
system and the Bohmian particle starts moving. However, in the
adiabatic limit its velocity is very small. Certainly, the Bohmian
particle does not visit every volume many times during the measurement
in such a way that the average time inside the volume $V$ is
proportional to $|\Psi (x)|^2$ -- a possible naive explanation based
on Bohm's ontology.

The breakdown of the second principle (as well as the first one) can be
seen much more clearly in an experiment suggested recently by Englert et
al. (1992). In order to explain their idea, we consider again our first
example in which the initial state of the particle is the superposition of
two identical wave packets moving in opposite directions; the Bohmian
particle initially is inside the right wave packet and we measure on the
left side the existence of the particle. Our measurement is robust in the
sense that the measuring device evolves into an orthogonal state if the
particle is in the left side, but the density of the wave function of the
measuring device is not changed significantly during the time of motion of
the particle. Englert et al. considered a spin flip which is a very clean
example, but, an example which fits better the spirit of the present work
is a standard von Neumann interaction which makes a shift of the wave
function of the pointer in the momentum representation. Then the
Hamiltonian is $ H = g(t) Q \Pi_{V}, $ instead of (1).  In fact, the
realization of the von Neumann measurement in the laboratory is of this
kind: in the Stern-Gerlach experiment the pointer is the particle itself
and it gets a shift in momentum due to the interaction. Only later, the
shift in momentum is translated into the shift in position. We consider the
case that the shift in momentum will lead to the shift in position
comparable with the width of the spatial wave function of the pointer only
much later, such that at the time of the overlap of the wave packets of the
particle, the two wave packets of the pointer, the one shifted in momentum
which correlated to the left wave packet of the particle and the other,
undisturbed, which correlated to the right wave packet, will essentially
coincide. In this case, the behavior of the Bohmian particle will be as in
the case that no measurement was performed at all. The Bohmian particle
will stop in the area of the overlap and will finally change the direction
of its motion, see Fig. 4. The particle started on the right will end up on the
right
side and will never reach the left side. The Bohmian particle of the
pointer, however, will follow the wave packet which got the shift in the
momentum: it eventually will move far away from its original place. This
corresponds to what we usually consider the result of the measurement
saying that the particle is on the left side. Therefore, it goes against
the first principle. But it also goes against the second principle. The
Bohmian particle was never on the left side. There was only an ``empty''
wave there. Nevertheless, it causes the action of the measuring device
located on the left side; the Bohmian particle of the pointer moves because
of the existence of the particle.

It is instructive to see how, exactly, the empty wave causes the
change in the measuring device. At the time of the interaction between
the particle and the measuring device, when the empty wave passes the
region $V$, no change in the motion of the Bohmian particle of the
pointer takes place. Only after the overlap of the empty and nonempty
waves of the particle,\break

\epsfysize=13 cm
 \centerline{ \epsfbox{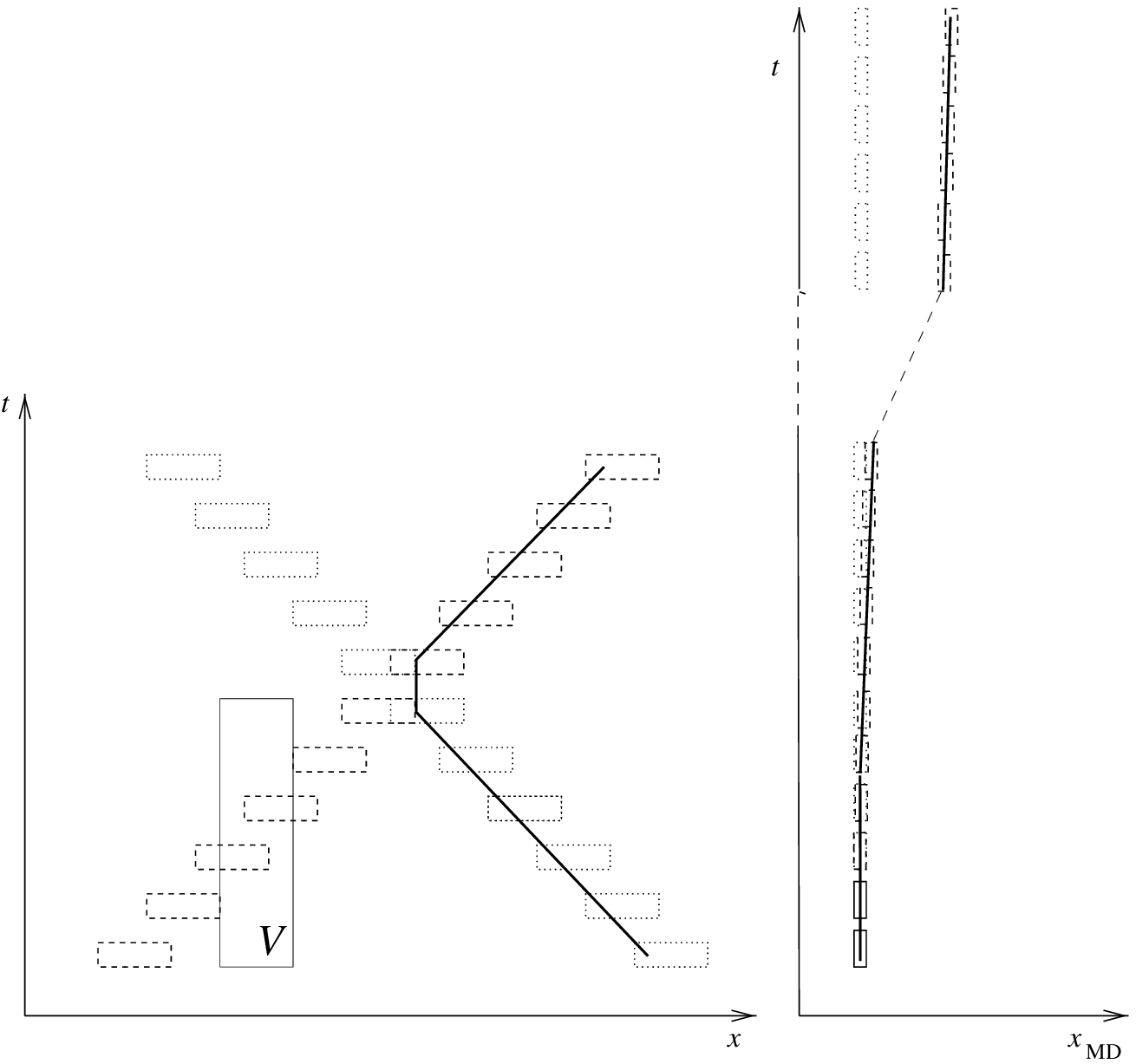}}
{\small {\bf Figure 4}: {\footnotesize
 Space-time diagram for the ``delayed measurement''.  The
Bohmian particle does not pass through the interaction region $V$,
but, nevertheless, the Bohmian particle of the measuring device moves.
Note that it starts moving not immediately after the interaction in
the region $V$, but only after the overlap of the two wave packets of
the particle.  The Bohmian particle of the pointer moves out of the
location of its initial wave packet only at much later time.
}}

\vspace {0.5 cm}

\noindent
the Bohmian particle of the pointer starts
moving. Note, that if we ``look'' on the problem in configuration
space, then no surprising behavior is observed. The empty wave in configuration
space does not influence the motion of the Bohmian
particle in configuration space until the particle is ``inside'' the
wave and the wave ceases to be empty. The difficulty is seen only in
the ordinary three dimensional space. It is even more dramatic when
our particle moves inside a special bubble chamber. The bubbles
created due to the passage of the particle are developed slowly enough
such that during the time of the motion of the particle the density of
the spatial wave function of each bubble does not change
significantly. Then, what we will see as a trace of bubbles is the
particle moving from the left to the right, while the Bohmian
particle, in fact, will move from the right side, stop, and come back
to the right. Note, however, that the picture we see via appearance of
the bubbles is time delayed. In a ``fast'' bubble chamber, in which we
see a real time motion, the Bohmian particle moves together with the
trace of bubbles.

The examples considered in this work do not show that the Bohm's causal
interpretation is inconsistent. It shows that Bohmian trajectories behave
not as we would expect from a classical type model. Before seeing the
examples discussed in this work we could think otherwise.  In fact we did,
and we were very reluctant to accept that the Bohmian picture is different, in
some cases, from a naive picture based on the outcomes of the experiments;
we worked hard, but in vain, searching for an error in our, and the Englert et
al., arguments.  If we follow David Bohm, viewing his theory as an
alternative formalism which should lead us to fruitful generalizations and
modifications, the difficulties we have discussed can be considered on the
positive side as showing a direction for constructing a better theory.
However, in order to put our work in a proper perspective, we have to note
that the proponents of the Bohm theory do not see the phenomena we
described here as difficulties of the theory; see, for example, D\"urr et
al. (1993).  The fact that we see these difficulties follows from our
particular approach to the Bohm theory in which the wave is not considered
to be a ``reality''. Thus, our difficulties are not surprising in the light
of the words of Bell (1987, p. 128): ``{\it No one can understand this
theory until he is willing to think of $\psi$ as a real objective field
rather than just a `probability amplitude'.}'' Although we are certainly
sympathetic to the approach in which $\psi$ is real (see Aharonov and
Vaidman 1993), one of us (L.V.) does not see it fit well into the framework
of the Bohm theory, since in his view, the main purpose of the Bohm theory
is to avoid the reality of the many worlds that he believes is incorporated in
the wave
function of the Universe (Vaidman 1993).

We are grateful to Sheldon Goldstein for illuminating correspondence which
clarified and sharpened enormously this work. This research was supported in
part by  grant 614/95 of the
Basic
Research Foundation (administered by the Israel Academy of Sciences and
Humanities)


\noindent

\vspace {.6cm}

\noindent
{\bf REFERENCES}

\vskip .32 cm \noindent
 Aharonov, Y., Albert, D., Casher, A., and Vaidman, L. (1987), ``Surprising
Quantum Effects,'' Physics Letters A124, 199-203.

\vskip .32 cm \noindent
 Aharonov, Y., Albert, D., and Vaidman, L. (1988), ``How the Result of
Measurement of a Component of the Spin of a Spin-1/2 Particle Can Turn Out
to Be 100,'' Physical Review Letters 60, 1351-54.

\vskip .32 cm \noindent
Aharonov, Y., Anandan, J., Popescu, S., and Vaidman, L.  (1990),
``Superpositions of Time Evolutions of a Quantum System and a Quantum Time
Machine,'' Physical Review Letters 64, 2965-68.

\vskip .32 cm \noindent
Aharonov, Y.,  Bergmann,  P.G., and  Lebowitz, J.L.(1964), ``Time Symmetry
in the Quantum Process of Measurement,'' Physical Review 134B, 1410-16.

\vskip .32 cm \noindent
Aharonov, Y.  and Vaidman, L.  (1990), ``Properties of a Quantum System
During the Time Interval Between Two Measurements,'' Physical Review A 41,
11-20.

\vskip .32 cm \noindent
Aharonov, Y.  and Vaidman, L. (1993), ``Measurement of the Schr\"odinger
Wave of a Single Particle,'' Physics Letters A178, 38-42.

\vskip .32 cm \noindent
Albert, D. (1992), Quantum Mechanics and Experience, Cambridge, Harvard
University Press.

\vskip .32 cm \noindent
 Bell, J. S. (1987), Speakable and Unspeakable in Quantum Mechanics.
Cambridge, Cambridge University Press.

\vskip .32 cm \noindent
 Bohm, D. (1952), ``A Suggested Interpretation of the Quantum Theory in
 Terms of `Hidden' Variables I and II,'' Physical Review 85, 97-117.

\vskip .32 cm \noindent
Dewdney, C., Holland P.R., Kyprianidids A., and Vigier J.P. (1988), ``Spin
and Non-Locality in Quantum Mechanics,'' Nature 336, pp. 536-544.

\vskip .32 cm \noindent
 D\"urr, D.,  Fusseder, W.,    Goldstein, S., and   Zanghi, Z. (1993),
``Comment on `Surrealistic Bohm Trajectories,''
 Zeitschrifft f\"ur Naturforschung  48a, 1261-62.

\vskip .32 cm \noindent
 Englert, B., Scully, M.O., S\"ussmann, G. and Walther, H. (1992)
``Surrealistic Bohm  trajectories,''
 Zeitschrifft f\"ur Naturforschung 47a, 1175-86.

\vskip .32 cm \noindent
Everett, H. (1957) ``Relative State' Formulation of Quantum Mechanics,''
Review of Modern Physics 29. 454-62.

\vskip .32 cm \noindent
Vaidman, L. (1993)
``About Schizophrenic Experiences of the Neutron or Why We
should Believe in the Many-Worlds Interpretation of Quantum Theory,''
Tel-Aviv University Preprint TAUP 2058-93.

\end{document}